\documentclass{ws-mpla}

\begin{document}

\markboth{GIORGIO PAPINI}
{Covariance and gauge invariance}

\catchline{}{}{}{}{}

\title{\bf COVARIANCE AND GAUGE INVARIANCE IN RELATIVISTIC THEORIES OF GRAVITY\\}

\author{\footnotesize GIORGIO PAPINI}

\address{Department of Physics and Prairie Particle Physics Institute, University of Regina, Regina, Sask S4S 0A2, Canada\\
and\\
International Institute for Advanced Scientific Studies,
89019 Vietri sul Mare (SA), Italy}


\maketitle

\pub{Received (Day Month Year)}{Revised (Day Month Year)}

\begin{abstract}
Any metric theory of gravity whose interaction with quantum
particles is described by a covariant wave equation is equivalent
to a vector theory that satisfies Maxwell-type equations
identically. This result does not depend on any particular set of
field equations for the metric tensor, but only on covariance. It
is derived in the linear case, but can be extended to any order of
approximation in the metric deviation. In this formulation
of the interaction of gravity with matter,
angular momentum and momentum are conserved
locally.

\keywords{Covariant wave equations; gravitational quantum phases; quantum gravity; maximal acceleration; strong gravity.}
\end{abstract}

\ccode{PACS Nos.04.62.+v: include PACS Nos.95.30.Sf}

\section{\bf Introduction. Solving the Klein-Gordon equation}	

Utiyama showed \cite{UTI} that important general
relativistic aspects of the interaction of gravity with matter
could be arrived at by generalizing the space-time independent
coordinate transformations of special relativity to local ones.
These results were later extended by Kibble \cite{KIB} using the
complete Poincar\'{e} group. More recently Capozziello and De Laurentis \cite{CAP}
have shown that general gauge theories of gravity can be derived
from local Poincar\'e symmetry. Unlike the approach followed in
\cite{UTI,KIB,CAP}, the present work derives a theory of the
interaction of gravity with particles described by covariant wave
equations by taking advantage of a restriction to global phase
invariance suggested by the covariant wave equations themselves.
The solution discussed in this Section does in fact contain
gravity in a space-time dependent phase which is the fundamental
ingredient of any gauge theory. The case of a particle described
by a Klein-Gordon equation is considered in detail, but similar
conclusions can be reached starting from other known wave
equations.

Covariant wave equations that apply to particles with, or without spin,
have solutions \cite{PAP0,PAP1,PAP2,PAP3,PAP4} that are exact to first order
in the metric deviation $\gamma_{\mu\nu}= g_{\mu\nu}-\eta_{\mu\nu}$, where $\eta_{\mu\nu}$ is the Minkowski metric,
and have been applied to problems like geometrical optics \cite{PAP3},
interferometry and gyroscopy \cite{PAP1}, the spin-flip of particles in gravitational and inertial fields \cite{PAP5},
radiative processes \cite{PAP6,PAP7} and spin currents \cite{PASP}.

It is useful to re-derive here the solution of the Klein-Gordon equation that, neglecting curvature dependent terms
and applying the Lanczos-De Donder condition
\begin{equation}\label{L}
  \gamma_{\alpha\nu,}^{\,\,\,\,\,\,\,
  \nu}-\frac{1}{2}\gamma_{\sigma,\alpha}^\sigma = 0 \,,
  \end{equation}
becomes to $\mathcal{O}(\gamma_{\mu\nu})$
\begin{equation}\label{KG}
\left(\nabla_{\mu}\nabla^{\mu}+m^2\right)\phi(x)\simeq\left[\eta_{\mu\nu}\partial^{\mu}\partial^{\nu}+m^2
+\gamma_{\mu\nu}\partial^{\mu}\partial^{\nu}
\right]\phi(x)=0\,.
\end{equation}
The notations and units ($\hbar=c=1$) are as in \cite{PAP5}. The
solution of (\ref{KG}) is obtained by solving the Volterra
equation
\begin{equation}\label{V}
\phi(x)=\phi_{0}(x)-\int_P^{x}d^4x' G(x,x') \gamma_{\mu\nu}(x')
\partial^{'\mu}\partial^{'\nu}\phi(x')\,,
\end{equation}
along the particle world-line, where $P$ is a fixed reference point, $x$ a generic point in
the physical future along the world-line, $G(x,x') $ is the causal Green function with
$(\partial^2 +m^2)G(x,x')=\delta^4 (x-x') $. The free Klein-Gordon equation is
\begin{equation}\label{KG0}
(\partial^2 +m^2)\phi_{0}=0\,.
\end{equation}
In first approximation
$\phi_{0}$ can be substituted for $\phi$ in (\ref{V}) and the integrations can then be carried out using the equations
\begin{equation}\label{E1}
\left(\partial_{x}^2+m^2\right)\frac{1}{2}\int_P^x dz^{\lambda}\gamma_{\alpha\lambda}(z)\partial^{\alpha}\phi_{0}(x)
=\gamma_{\alpha\beta}(x)\partial^{\alpha}\partial^{\beta}\phi_{0}(x)+
\frac{1}{2}\gamma_{\alpha\mu,}^{\,\,\,\,\,\,\,\mu}(x)\partial^{\alpha}\phi_{0}(x)\,,
\end{equation}
from which $\gamma_{\alpha\beta}(x)\partial^{\alpha}\partial^{\beta}\phi_{0}(x)$ can be obtained, and
\begin{equation}\label{E2}
\left(\partial_{x}^2 +m^2\right)\frac{1}{2}\int_P^x dz^{\alpha}\int_P^x dz^{\lambda}\left(\gamma_{\alpha\lambda,\beta}(z)
-\gamma_{\beta\lambda,\alpha}(z)\right)\partial^{\beta}\phi_{0}(x)
\end{equation}
\[=\left(\partial_x^2 +m^2\right)\frac{1}{2}\int_P^x dz^{\lambda}\left(\gamma_{\alpha\lambda,\beta}(z)
-\gamma_{\beta\lambda,\alpha}(z)\right)\left(x^{\alpha}-z^{\alpha}\right)\partial^{\beta}\phi_{0}(x)\]
\[=
\frac{1}{2}\left(\gamma^{\mu}_{\,\,\,\mu,\beta}(x)-\gamma_{\beta\mu,}^{\,\,\,\,\,\,\,\mu}(x)\right)
\partial^{\beta}\phi_{0}(x)\,.\]
Equations (\ref{E1}) and (\ref{E2}) can be proven by straightforward differentiation.
The latter is the four-dimensional extension of a known formula \cite{VER}.
To $\mathcal{O(\gamma_{\mu\nu})}$ the solution
of (\ref{KG}) is
\begin{equation}\label{SO}
\phi(x)=\left(1-i\hat{\Phi}_{G}(x)\right)\phi_{0}(x)\,,
\end{equation}
where the operator $\hat{\Phi}_G $ is defined as \cite{PAP5}
\begin{equation}\label{PHI}
\hat{\Phi}_{G}(x)=-\frac{1}{2}\int_P^x
dz^{\lambda}\left(\gamma_{\alpha\lambda,\beta}(z)-\gamma_{\beta\lambda,\alpha}(z)\right)
\left(x^{\alpha}-z^{\alpha}\right)\hat{k}^{\beta}
\end{equation}
\[+\frac{1}{2}\int_P^x dz^{\lambda}\gamma_{\alpha\lambda}\hat{k}^{\alpha}\,,\]
$\hat{k}^{\alpha}=i\partial^{\alpha}$ and $k^{\alpha}$ is the
momentum of the plane wave solution $\phi_0$ of (\ref{KG0})
satisfying $k_{\alpha}k^{\alpha}=m^2$. The solution {\it is
independent of any field equations for $\gamma_{\mu\nu}$}.
Equations (\ref{SO}) and (\ref{PHI}) are the byproduct of
covariance (minimal coupling) and, ultimately, of Lorentz
invariance and can therefore be applied to general relativity, in
particular to theories in which acceleration has an upper limit
\cite{CAI1,CAI2,CAI3,BRA,MASH1,MASH2,MASH3,TOLL} and that therefore allow the resolution of
astrophysical \cite{SCHW,RN,KERR,PU} and cosmological singularities
in quantum theories of gravity \cite{ROV,BRU}. They also
are  relevant to those theories of asymptotically safe gravity
that can be expressed as Einstein gravity coupled to a scalar
field \cite{CA}.

The calculation of $\phi$ can be extended to any order in
$\gamma_{\mu\nu}$, but the results cannot be summed up in closed
form. In fact, the solution can be  written in the form $\phi
=\Sigma_h \phi_{(h)}$ where \cite{PAP1}
\begin{equation}\label{phih}
\phi_{(h)}(x)=-i\hat{\Phi}_G\phi_{(h-1)}(x)
\end{equation}
\[=\frac{1}{2}\int_P ^x
dz^{\lambda}\left[\left(\gamma_{\alpha\lambda,\beta}-\gamma_{\beta\lambda,\alpha}\right)
\left(x^{\alpha}-z^{\alpha}\right)\partial^{\beta}-\gamma_{\alpha\lambda}\partial^{\alpha}\right]\phi_{(h-1)}(x)\,.\]
Any possible non-linearities present in the wave equation can be
treated as perturbations, where applicable \cite{Des}. The results
also apply, with appropriate changes, to all known covariant wave
equations \cite{PAP0}.

\section{\bf Maxwell-type equations}

By substituting (\ref{SO}) and (\ref{PHI}) in (\ref{KG}) one finds
that $\nabla_{\mu}\phi\rightarrow
(\nabla_{\mu}-i\Phi_{G,\mu})\phi$, where
\begin{equation}\label{PM}
\Phi_{G,\mu}(x)=-\frac{1}{2}\int_{P}^x dz^{\lambda}
\left(\gamma_{\mu\lambda,\alpha}-\gamma_{\alpha\lambda,\mu}\right)k^{\alpha}+\frac{1}{2}\gamma_{\alpha\mu}(x)
k^{\alpha}\,.
\end{equation}
It follows that the particle momentum is
$p_{\mu}=k_{\mu}-\Phi_{G,\mu}$. By using (\ref{PM}), one can also
write
\begin{equation}\label{nabla}
\nabla^{\mu}\rightarrow
\nabla^{\mu}-i\partial_{x}^{\mu}\left\{-\frac{1}{2}\int_P^xdz^{\lambda}
\left[\left(\gamma_{\alpha\lambda,\beta}(z)-
\gamma_{\beta\lambda,\alpha}(z)\right)\left(x^{\alpha}-z^{\alpha}\right)-
\gamma_{\beta\lambda}(z)\right]k^{\beta}\right\}
\end{equation}
\[= \nabla^{\mu}-iK^{\mu}(z,x)\equiv \mathcal{D}^{\mu}\,,\]
where $ \Phi_{G}(x)=\int_P^x dz^{\lambda}K_{\lambda}(z,x)$ and
\begin{equation}\label{K}
K_{\lambda}(z,x)=-\frac{1}{2}\left[\left(\gamma_{\alpha\lambda,\beta}(z)
-\gamma_{\beta\lambda,\alpha}(z)\right)\left(x^{\alpha}-z^{\alpha}\right)-\gamma_{\beta\lambda}(z)\right]k^{\beta}\,.
\end{equation}
In the form (\ref{nabla}) of the covariant derivative, the
two-point vector $K_\lambda(z,x)$ plays the role of the vector
potential in electromagnetism.

The definition of $K_\lambda$ given by (\ref{K}) contains a
reference to matter through the momentum $k_\mu$ of $\phi_0$. The
inclusion of the coupling to matter in $K_\lambda$ does not
however create problems in a theory of particle-gravity
interactions and offers the benefits of dealing with completely
gauge invariant quantities (Section 4).

The equations satisfied by $K_\lambda$ relative to the base point
$x^\alpha$ can now be obtained by differentiating (\ref{K}) with
respect to $z$. One finds
\begin{equation}\label{F}
F_{\mu\lambda}(z,x)=K_{\lambda,\mu}(z,x)-K_{\mu,\lambda}(z,x)=R_{\alpha\beta\lambda\mu}(z)\left(x^{\alpha}
-z^{\alpha}\right)k^{\beta}=R_{\mu\lambda\alpha\beta}(z) J^{\alpha\beta}\,,
\end{equation}
where
$R_{\alpha\beta\lambda\mu}(z)=\frac{1}{2}\left(\gamma_{\alpha\lambda,\beta\mu}
+\gamma_{\beta\mu,\alpha\lambda}-\gamma_{\alpha\mu,\beta\lambda}-\gamma_{\beta\lambda,\alpha\mu}\right)$
is the linearized Riemann tensor satisfying the identity
$R_{\mu\nu\sigma\tau}+R_{\nu\sigma\mu\tau}+R_{\sigma\mu\nu\tau}=0$
and
$J^{\alpha\beta}=\frac{1}{2}\left[\left(x^{\alpha}-z^{\alpha}\right)k^\beta-k^\alpha
\left(x^\beta-z^\beta\right)\right]$ is the angular momentum about
the base point $x^\alpha$. The physical origin of the base event
in the two-point tensors (\ref{K}) and (\ref{F}) is closely linked
to $J^{\alpha\beta}$. The latter tensor would of course contain
also the intrinsic angular momentum when referring to particles
with spin. Maxwell-type equations
\begin{equation}\label{ME1}
F_{\mu\lambda,\sigma}+F_{\lambda\sigma,\mu}+F_{\sigma\mu,\lambda}=0
\end{equation}
and
\begin{equation}\label{ME2}
F^{\mu\lambda}_{\,\,\,\,\,\,\,,\lambda}\equiv j^{\mu}=
\left(R^{\mu\lambda}_{\,\,\,\,\,\,\,\alpha\beta}J^{\alpha\beta}\right),_{\lambda}
=R^{\mu\lambda}_{\,\,\,\,\,\,\,\alpha\beta,\lambda}\left(x^\alpha
-z^\alpha\right)k^\beta +R^{\mu}_{\,\,\,\,\beta}k^{\beta}\,,
\end{equation}
can be obtained from (\ref{F}) using the Bianchi identities $R_{\mu\nu\sigma\tau,\rho}
+R_{\mu\nu\tau\rho,\sigma}+R_{\mu\nu\rho\sigma,\tau}=0$. The current $j^{\mu}$ satisfies the conservation law
$j^{\mu}_{\,\,\,,\mu}=0$.
One finds in particular that the "electric" and "magnetic" components of $F_{\mu\nu}$ are
\begin{equation}\label{COM}
E_{i}=R_{0i\alpha\beta}J^{\alpha\beta}\,\,,
H_{i}=\epsilon_{ijk}R^{kj}_{\,\,\,\,\,\,\,\alpha\beta}J^{\alpha\beta}\,.
\end{equation}
Equations (\ref{ME1}) and (\ref{ME2}) are identities and do not represent additional constraints on $\gamma_{\mu\nu}$.
The recombination of ten $\gamma_{\mu\nu}$ into four $K_{\lambda}$ may be regarded as an example of hidden
symmetry made manifest by the covariance of the wave equation, hence the interaction of the particle with
$\gamma_{\mu\nu}$. Knowledge of $\gamma_{\mu\nu}$ is still needed in order to calculate $K_\lambda$ and that
requires the solution of ten equations. One needs in fact the fifteen components $R_{0i23}$, $R_{0i13}$,
$R_{0i12}$, $R_{0i02}$,
$R_{0101}$, $R_{0103}$, $R_{0203}$ to determine $F_{0i}$ and the six components $R_{1212}, R_{1213}, R_{1313},
R_{1223}, R_{1323}, R_{2323}$ to calculate $F_{ij}$. The sum must be decreased by one because $R_{0123}+R_{0231}
+R_{0312}=0$ and the total, twenty, corresponds to the number of independent components of the Riemann tensor.
In general, therefore, all $\gamma_{\mu\nu}$ are required in order to determine $F_{\mu\nu}$.

The somewhat inverse approach, whereby gravitons are constructed as photon bound states, is discussed in \cite{PAPBS}.

\section{\bf Conservation laws}

Conservation laws are better expressed geometrically in integral form. Several of them are derived in a purely
classical context in \cite{Synge} using world function and two-point tensor formalisms. It is shown in particular
that for any skew-symmetric tensor $S_{\mu\nu}$ and any closed two-space $V_{2}$ in space-time spanned by an open
$V_3$ one has
\begin{equation}\label{SY}
\oint_{V_{2}}S_{\mu\nu} d\tau^{\mu\nu\sigma}=\int_{V_{3}}S_{\mu\nu,\sigma}d\tau^{\mu\nu\sigma}
=\frac{1}{3}\int_{V_{3}}\left(S_{\mu\nu,\sigma}+S_{\nu\sigma,\mu}+S_{\sigma\mu,\nu}\right)d\tau^{\mu\nu\sigma}\,.
\end{equation}
On substituting $F_{\mu\nu}$ for $S_{\mu\nu}$ and taking (\ref{ME1}) into account, (\ref{SY}) gives
\begin{equation}\label{CON1}
\oint_{V_{2}}F_{\mu\nu}d\tau^{\mu\nu}=\oint_{V_{2}}R_{\mu\nu\alpha\beta}J^{\alpha\beta}d\tau^{\mu\nu}=0\,,
\end{equation}
which states that the flux of $J^{\alpha\beta}$ relative to a base point $x^\alpha$ out of a closed surface vanishes.
The same result can be obtained by integrating  $K_{\lambda}$ over a closed $V_{1}$ spanned by an open $V_2$
and letting $V_{1}$ shrink to a point. It also follows that, when the wave function is nodal, the flux of $K_{\lambda}$
\begin{equation}\label{FQ}
\oint_{V_1}dz^{\lambda}
K_{\lambda}(z,x)=\int_{V_2}R_{\mu\nu\alpha\beta}J^{\alpha\beta}d\tau^{\mu\nu}\,,
\end{equation}
is quantized.
Finally, the same procedure applied to $j_\mu$ gives
\begin{equation}\label{j}
\oint_{V_1}j_{\mu}dz^\mu =\int_{V_2} j_{\mu,\nu}d\tau^{\mu\nu}=\int_{V_2}
\left(R_{\alpha\beta\mu}^{\,\,\,\,\,\,\,\,\,\,\,\lambda}J^{\alpha\beta}\right)_{,\lambda\nu}d\tau^{\mu\nu}
\end{equation}
and, by contracting $V_1$ to a point and using Bianchi identities,
\begin{equation}\label{j1}
\oint_{V_2}\left(R_{\alpha\beta\mu}^{\,\,\,\,\,\,\,\,\,\,\,\lambda}J^{\alpha\beta}\right)_{,\lambda\nu}d\tau^{\mu\nu}
=\oint_{V_2}\left[\left(R_{\mu\alpha,\beta\nu}-R_{\mu\beta,\alpha\nu}\right]J^{\alpha\beta}
+2R_{\mu\beta,\nu}k^{\beta}\right]d\tau^{\mu\nu}=0\,,
\end{equation}
that again states that the total flux of angular momentum and
momentum out of a closed $V_2$ vanishes. Since $V_2$ can be chosen
arbitrarily small and $j^{\mu}$ is conserved, one concludes that
the total flux of momentum and angular momentum is conserved
locally. This is shown in \cite{CAP} to be a necessary ingredient of
gauge theories of gravity. It appears here in a context still
independent of any choice of gravitational field equations.
Local conservation requires the introduction of
the field
$K_\lambda$. The corresponding force field is represented by
(\ref{F}) that contains curvature in an essential way.

\section{\bf Gauge transformations and equations of motion}

There are two types of "gauge" transformations at play in what
follows. First are those that follow from (\ref{F}) and leave
$F_{\mu\nu}$ invariant under the changes
\begin{equation}\label{GT}
K'_{\mu}(z,x)=K_{\mu}(z,x)-\Lambda_{,\mu}(z)\,.
\end{equation}
It also follows from (\ref{nabla}) and (\ref{GT}) that
$e^{i\Lambda(z)}\mathcal{D}_{\mu}e^{-i\Lambda(z)}=\partial_{\mu}-iK'_{\mu}(z,x)=\mathcal{D}'_{\mu}$.
These are the usual gauge transformations one must expect in a
theory formally analogous to electromagnetism with symmetry group
$U(1)$. They can also be used to eliminate redundant terms from
the equations, as follows. Differentiating (\ref{K}) twice with
respect to $z^{\mu}$ one finds
\begin{equation}\label{X}
\partial^2 K_{\lambda}=-\frac{1}{2}\left[\left(\partial^2 \gamma_{\alpha\lambda,\beta}-\partial^2
\gamma_{\beta\lambda,\alpha}\right)\left(x^{\alpha}-z^{\alpha}\right)+\partial^2 \gamma_{\beta\lambda}-\partial^2
\gamma_{,\lambda}\right]k^{\beta}\,,
\end{equation}
where $\partial^2=\partial_{\mu}\partial^{\mu}$ and $\gamma=\gamma_{\mu}^{\,\,\,\,\mu}$. The last term can be
eliminated from (\ref{X}) by using (\ref{GT}) with $\Lambda=-\frac{1}{2}\gamma_{,\beta}k^{\beta}$.
The corresponding equation for $F_{\lambda\mu}$ can be obtained by differentiating (\ref{F}) and using (\ref{L})
\begin{equation}\label{Y}
\partial^2 F_{\lambda\mu}=\frac{1}{4}\left[\partial^2\left(-\eta_{\alpha\mu}\gamma_{,\beta\lambda}
-\eta_{\beta\lambda}\gamma_{,\alpha\mu}+\eta_{\alpha\lambda}\gamma_{\beta\mu}+\eta_{\beta\mu}
\gamma_{,\alpha\lambda}\right)\right]\left(x^{\alpha}-z^{\alpha}\right)k^{\beta}
\end{equation}
\[-\frac{1}{2}\partial^{2}\left(-\eta_{\beta\lambda}\gamma_{,\mu}+\eta_{\beta\mu}\gamma_{,\lambda\beta}\right)
k^{\beta}\,,\] and is invariant under (\ref{GT}). The second
instance in which the term "gauge transformation" is traditionally
used is related to the fact that (\ref{L}) does not determine the
coordinates completely. The additional, allowable transformations
$z^{\alpha}\rightarrow z^{\alpha}+\xi^{\alpha}$, where
$\xi^{\alpha}$ are first order quantities, induce the changes
\begin{equation}\label{gg}
\gamma_{\alpha\lambda}\rightarrow
\gamma_{\alpha\lambda}-\xi_{\alpha,\lambda}-\xi_{\lambda,\alpha}\,.
 \end{equation}
Because $K_\lambda$ contains $\gamma_{\mu\nu}$ explicitly, one
must ascertain that $K_\lambda$ and the equations it satisfies
behave in a way that is consistent with (\ref{gg}) and (\ref{GT}).
One finds that (\ref{gg}) still leaves (\ref{X}) invariant
provided $\partial^2 \xi_{\alpha}=0$ and affects $K_\lambda$ in a
way consistent with (\ref{GT}). In fact, by applying (\ref{gg}) to
$K_\lambda$ one obtains
$K'_{\lambda}=K_\lambda-\partial_\lambda\left\{\frac{1}{2}\left(-\xi_{\alpha,\beta}
+\xi_{\beta,\alpha}\right)\left(x^{\alpha}-
z^\alpha\right)k^{\beta}\right\}$ as required by (\ref{F}).

Bel suggested \cite{Bel} an alternative way to strengthen the
analogy between electromagnetism and linearized gravity by
introducing the tensor
$F_{\alpha\beta\lambda}=\gamma_{\beta\lambda,\alpha}-\gamma_{\alpha\lambda,\beta}$
that also satisfies Maxwell-type equations. However
$F_{\alpha\beta\lambda}$ transforms under (\ref{gg}) as
$F_{\alpha\beta\lambda}\rightarrow
F_{\alpha\beta\lambda}+\partial_{\lambda}\left(\xi_{\beta,\alpha}-\xi_{\alpha,\beta}\right)$
and is not therefore a gauge invariant quantity. Interesting
results can nonetheless be obtained by restricting (\ref{gg}) to
the special functions $\xi_{\mu}=\partial_{\mu}\xi$ \cite{Bel}.
This additional restriction is, of course, unnecessary when
dealing with $K_\lambda$ and $F_{\alpha\beta}$.

Suppose now that
$\gamma_{\mu\nu}=\varepsilon_{\mu\nu}\exp(ip_{\alpha}z^{\alpha})$
and that the wave propagates in the direction characterized by the
wave vector $p^1=p^2=0$ and $p^3=p^0>0$. Then (\ref{L}) and
(\ref{gg}) provide, among the surviving components, the relations
$\varepsilon_{22} =,\varepsilon_{11},
\varepsilon_{02}=-\varepsilon_{32},\varepsilon_{01}=-\varepsilon_{31},
\varepsilon_{03} =-\frac{1}{2}(\varepsilon_{33}+\varepsilon_{00})$
\cite{WEI}. One finds
\begin{equation}\label{PO1}
K_1 =-\frac{1}{2}\gamma_{01}\left\{k^0\left[\left(-ip_0 -ip_3\right)\left(x^3 -z^3\right)-1\right]
-k^3\left[\left(ip_3 +ip_0\right)\left(x^0 -z^0\right)-1\right]\right\}
\end{equation}
\[K_2 = -\frac{1}{2}\gamma_{02}\left\{k^0\left[\left(-ip_0 -ip_3\right)\left(x^3 -z^3\right)-1\right]
-k^3\left[\left(ip_3 +ip_0\right)\left(x^0 -z^0\right)-1\right]\right\}\,.\]
 It then follows from $K_1\pm iK_2$ that the helicity is $\pm 1$. On the contrary $K_0, K_3$ depend only
 on $\varepsilon_{00}$ and $\varepsilon_{33}$ with helicity zero.

Pursuing the "electromagnetic" analogy, one can also derive the equations of motion of a classical particle.
On account of (\ref{F}) and writing $\eta^{\alpha}\equiv x^{\alpha}-z^{\alpha}$, one finds
\begin{equation}\label{EM}
\frac{d^2\eta^{\mu}}{ds^2}=F^{\mu}_{\,\,\,\,\lambda}u^{\lambda}\equiv
R^{\mu}_{\,\,\,\,\beta\lambda\alpha}\eta^{\alpha}u^{\beta}u^{\lambda}\,,
\end{equation}
which is the equation of geodesic deviation. Similarly, the equation that relates the torque to the change in
angular momentum is
\begin{equation}\label{AM}
\eta_{\mu\alpha\beta\rho}\frac{du^{\alpha}}{ds}\eta^{\beta}
=\eta_{\mu\alpha\beta\rho}F^{\alpha}_{\,\,\,\,\,\tau}u^{\tau}\eta^{\beta}\equiv\eta_{\mu\alpha\beta\rho}\eta^{\beta}
\eta^{\omega}R^{\alpha}_{\,\,\,\,\,\sigma\omega\tau}u^{\sigma}u^{\tau}\,.
\end{equation}

\section{\bf Conclusions}

 The main results are summarized below.

Any linear theory of gravity whose interaction with quantum
particles is described by a covariant wave equation leads to a
vector theory in which angular momentum and momentum are conserved
locally. The ten components of $\gamma_{\mu\nu}(x)$ regroup to
form a two-point vector $K_{\mu}(z,x)$ that satisfies Maxwell-type
equations identically. {\it The result is independent of the choice of
any particular set of field equations for the metric tensor} and
therefore applies to a variety of theories of gravity, in
particular to general relativity and to Caianiello's theory of
maximal acceleration. In general, knowledge of $K_{\mu}(z,x)$
still requires knowledge of all components of
$\gamma_{\mu\nu}(x)$. The construction of $K_{\mu}$ can be
extended to all orders of approximation in the metric deviation.

The field $K_\lambda$ renders the local conservation
of angular momentum and momentum possible. This is borne out of
the integral laws (\ref{CON1}) and (\ref{j1}) that state that the
flux of $J_{\alpha\beta}$ across a closed $V_2$ is conserved. This is a
general feature of gauge theories of gravity that follows \cite{CAP}
from local Poincar\'e symmetry. If,
in particular, the particle wave function is nodal, then the flux
of $K_{\lambda}$ out of an open $V_{2}$ is quantized. 

The matter considered
is a spinless particle. The introduction of spin \cite{PAP2,PAP3,PAP4,PAP5}
necessitates that
$J^{\alpha\beta}$ refer to the total angular momentum in
agreement with \cite{CAP}.

All results are invariant under the transformations (\ref{GT}) and
(\ref{gg}) and cannot therefore be an artifact of the choice of
coordinates. In the case $\gamma_{\mu\nu}(x)$ represents a
gravitational wave, the helicity of $K_{\mu}(z,x)$ is that
expected of a (not necessarily free) vector field.

The "electromagnetic" analogy can also be pursued at the level of the particle equations of motion.
It yields in this instance the equation of geodesic deviation  and a generalized torque-change of
angular momentum relation.

\end{document}